# Double-bowl State in photonic Dirac nodal line semimetal

Mengying Hu[1], Ye Zhang[1], Xi Jiang[1], Tong Qiao[1], Qiang Wang[2], Shining Zhu[1], Meng Xiao[3], and Hui Liu[1]

[1]*National Laboratory of Solid State Microstructures, School of Physics, Collaborative Innovation Center of Advanced Microstructures, Nanjing University, Nanjing 210093, China.*

[2]*Division of Physics and Applied Physics, School of Physical and Mathematical Sciences, Nanyang Technological University, Singapore 637371, Singapore.*

[3]*Key Laboratory of Artificial Micro- and Nano-structures of Ministry of Education and School of Physics and Technology, Wuhan University, Wuhan 430072, China.*

Correspondence: Meng Xiao (phmxiao@whu.edu.cn); Hui Liu (liuhui@nju.edu.cn).

## Abstract

The past decade has seen a proliferation of topological materials for both insulators and semimetals in electronic systems and classical waves. Topological semimetals exhibit topologically protected band degeneracies, such as nodal points and nodal lines. Dirac nodal line semimetals (DNLS), which own four-fold line degeneracy, have drawn particular attention. DNLSs have been studied in electronic systems but there is no photonic DNLS. Here in this work, we provide a new mechanism which is unique for photonic systems to investigate a stringent photonic DNLS. When truncated, the photonic DNLS exhibits double-bowl states (DBS), which comprises two sets of perpendicularly polarized surface states. In sharp contrast to nondegenerate surface states in other photonic systems, here the two sets of surface states are almost degenerate over the whole spectrum range. The DBS and the bulk Dirac nodal ring (DNR) dispersion along the relevant directions, are experimentally resolved.

## Introduction

Discovering new topological phases of matter is of significant importance for both fundamental physics and materials science[1-7]. Theory of symmetry indicators is successful in identifying electronic topological materials[8]. With mature algorithm developed, extensive efforts have been taken to diagnose topological characters of electronic materials in the crystal structure database exhaustively[9-11]. The topological classification of the photonic systems was originally thought to be a trivial extension of the electronic counterpart and described by spinless space groups. However, detailed analyses reveal that photonic systems are distinct from the electronic counterparts, and connectivity at zero frequency in dielectric materials and hidden symmetry enforced nexus points are latter found to be unique to photonic systems[12,13]. Here in this work, we provide a stringent photonic realization of Dirac nodal line semimetal (DNLS) which is not a spinless version of the electronic DNLS. More intriguingly, such a photonic DNLS exhibits perpendicularly polarized double-bowl surface states (DBS), which are degenerately pinned at the bowl center and bowl edge and are almost degenerate over the entire spectrum range. This is in sharp contrast to other photonic systems, where the two perpendicularly polarized states are in general nondegenerate.

DNLSs[14-18] and three-dimensional (3D) Dirac semimetals[19] with four-fold band degeneracy stand as important members of the topological semimetal family[14-24]. They exhibit various unique properties such as giant diamagnetism[25], flat Landau levels[26], and long-range Coulomb interaction[27], among others[28]. In addition, they are neighbors to many novel topological phases and thus serve as ideal platforms for investigating topological phase transitions[19]. Three-dimensional Dirac semimetals have been observed in both electronic systems and classical waves[19,29-31]. In electronic systems, DNLSs are possible in the absence of spin-orbital couplings[16-18]. Meanwhile, they can also be protected by nonsymmorphic symmetries in the presence of spin-orbital couplings[14,15]. However, there is NO photonic DNLS in all previous works.

The effective Hamiltonian of a simple Dirac nodal ring (DNR) degeneracy of a DNLS in the *x-y* plane can be written:

$$H = \left[\left(q_\rho - q_0\right)\sigma_x + q_z\sigma_z\right]\tau_0 \qquad (1)$$

where $\tau_0$ is the $2\times 2$ identity matrix, $\sigma_x$ and $\sigma_z$ are Pauli matrixes, $q_0$ is the radius of the nodal ring, and $q_\rho$ and $q_z$ represent the wave vector along the radial and $z$ directions, respectively. Such a Hamiltonian possesses a four-fold ring degeneracy along the polar angle $\hat{\varphi}$ direction at $q_\rho = q_0$ and $q_z = 0$. According to convention, we use a $\sigma$ matrix to represent the band index and a $\tau$ matrix to represent the (pseudo-) spin index. A two-fold Weyl nodal line degeneracy can be easily constructed by intersecting two bands with different representations of a certain symmetry such as the mirror symmetry, PT symmetry, or glide symmetry[14]. However, the extension of the two-fold Weyl line degeneracy to a four-fold degeneracy DNR is not an easy task. Currently, all the nodal line semimetals in classical waves have been Weyl nodal line semimetals with two-fold band degeneracy[24,32-34], and there has been no DNLS in classical waves. For light, the pseudo-spin freedom is interpreted as the polarizations. In principle, the optical responses of different polarizations can be tuned to be identical by setting $\varepsilon = \mu$; however, this is experimentally impractical. Additionally, the lattice symmetries exhibit higher dimensional representations only at highly symmetric points. The nonsymmorphic symmetries that protect the electronic DNR do not work for light due to the inherent distinction between fermions (electron) and bosons (photon); *i.e.*, the time reversal operator squares to –1 for fermions, whereas it is +1 for bosons. Thus, to construct an optical DNLS, a brand new mechanism is here established in order to ensure that the coefficients in front of all $\tau_{x,y,z}$ matrixes in the Hamiltonian vanish over a certain parameter range.

## Results

Our system is an AB layered photonic crystal (PC) with $SiO_2$ ($\varepsilon_A \approx 2.18$) and $d_A = 388$nm for layer A, and $Ta_2O_5$ ($\varepsilon_B \approx 5.06$) and $d_B = 597$nm for layer B. This structure can be fabricated with the e-beam evaporation technique. An SEM picture of our sample is shown in Fig. 1a. Simple as it is, this structure is perceived as a photonic DNLS. The band degeneracies of this system are only found at $k_z = 0$ or $k_z = \pi/\Lambda$, (see Ref. [35] and Supplementary information Sec. I), where $\Lambda = d_A + d_B$ is the unit cell length and $k_z$ is the Bloch wave vector perpendicular to the layers. Thereby, to identify the band degeneracies of this system, it is only necessary to plot the band edge states at at $k_z = 0$ and $k_z = \pi/\Lambda$ as a function of $k_x$ while keeping $k_y = 0$, as shown in Fig. 1b. The pseudo-spin freedom shown in the figure consists of the transverse-electric (TE) and transverse-magnetic (TM) modes, featured by the electric and magnetic fields only in the in-plane directions, respectively. Basically, there are two types of band degeneracies, marked in Fig. 1b with black circles and orange circles. The black circle degeneracies are due to the Brewster angle where the impedances of the two layers match, which is only possible for the TM modes (the red and green bands). They lie along an essentially straight black dashed line since the material dispersions are small in the frequency range of interest (index variation < 1.9% for $SiO_2$ and < 6.4% for $Ta_2O_5$; see Supplementary Data I for measured refractive index).

In addition to the degeneracy induced by the Brewster angle, there is another type of band degeneracies for both the TE and TM modes, marked with orange circles in Fig. 1b. Interestingly, the degeneracies of TE and TM modes occur at identical $k_x$s. It is here proven that (see proof in Supplementary information Sec. I) such band degeneracies exist when

$$\tilde{n}_A d_A / \tilde{n}_B d_B = m_1 / m_2 \in \mathbb{Q} \tag{2}$$

where $m_1, m_2 \in \mathbb{N}^+$, and $\tilde{n}_i = \sqrt{\varepsilon_i \mu_i - k_x^2 / k_0^2}$ ($i = A$ or $B$) with $k_0$ being the wave vector in a vacuum. Under such a condition, the $(m_1 + m_2)^{th}$ band and the $(m_1 + m_2 + 1)^{th}$ band cross at

$$f_{m_1+m_2} = (m_1 + m_2) c / 2(\tilde{n}_A d_A + \tilde{n}_B d_B) \tag{3}$$

where $c$ is the speed of light in a vacuum. Equations (2) and (3) extend the relationship in Ref. [33] at the normal direction to off-normal directions at finite $k_x$. We emphasize that Eqs. (2) and (3) work for both TE and TM polarizations and hence lead to a four-fold degeneracy. In addition, the existence of such band degeneracies is not accidental. It depends little on the material dispersions nor requires specific materials. Moreover, $k_x$s of the four-fold degeneracies can be simply controlled by varying $d_A$ and $d_B$ for the chosen materials (see more details in Supplementary information Sec. I).

We here focus on one of the four-fold degeneracies around 591THz, as stressed in Fig. 1b. Considering the fact that our system is rotationally invariant, if there exists a degeneracy at $k_x = k_{\rho D}$ and $k_y = 0$, such a degeneracy ought to be extended to form a ring shape at $k_{\rho D} \equiv \sqrt{k_x^2 + k_y^2}$. Figure 1c sketches the in-plane band degeneracy around $k_{\rho D} = 1.32(2\pi/\Lambda)$, from which we can see that two linear TM (red) bands are sandwiched by two linear TE (blue) bands and all the bands are degenerated at the golden ring at $k_{\rho D}$. The dispersions of these four bands are all positive along the in-plane radial direction. Figure 1d shows the dispersion along the $k_z$ direction around this four-fold degeneracy. All the bands have linear dispersion away from the degenerate point. Thus, we evidence the existence of DNLS with nodal line degeneracy at $k_{\rho D} = 1.32(2\pi/\Lambda)$, $k_z = \pi/\Lambda$, and $f = 591 \text{THz}$, as sketched in Fig. 1e. Combined with the dispersions shown in Figs. 1b and 1d, one can conclude that the DNLS shown in Fig. 1 belongs to type II[36]. The ring degeneracy for each polarization in our system is protected by the intrinsic mirror symmetry of the AB layered structure. Meanwhile, the degeneracy between different polarizations is required by Eqs. (2) and (3). In contrast to the DNLS protected by the nonsymmorphic symmetries in electronic systems, in our system, the DNRs can be found at both $k_z = 0$ and $k_z = \pi/\Lambda$. (The DNR at $k_z = 0$ is provided in Supplementary information Sec. II.)

The experimental setup is depicted schematically in Fig. 1f, where we use yellow and brown to depict the layers made of $SiO_2$ and $Ta_2O_5$, respectively. The number of unit cells of our PC is 12, which is not shown explicitly in Fig. 1f. Light is incident (magenta) with either the TE or TM polarization on the sample with the incident angle $\theta$, which determines the parallel wave vector excitation. In our experiments, we cover $\theta$ from $0°$ to $60°$. The azimuth angle $\varphi$ can be flexibly tuned to verify the in-plane isotropy. The transmission and reflection spectra as functions of the frequency ($f$), $\theta$ and $\varphi$ are here collected and analyzed.

We perform angle-resolved transmission measurements on the PC sample to probe the dispersions of the DNLS, in which we start by setting $\varphi = 0°$ without loss of generality. The experimental results are given in Figs. 2a and 2b for the TE and TM polarizations, respectively, alongside with the corresponding simulation spectra in Figs. 2c and 2d for comparison. We then carry out measurements at other $\varphi$s (Supplementary information Sec. III), and the measurements are almost identical to those at $\varphi = 0°$, which thus fully confirms the in-plane isotropy of our sample. Notice that the numerical and experimental results are closely consistent with each other. Since the dispersions along the $k_z$ direction are monotonic (Supplementary information Sec. I), the transmission spectra fill the projected band region with the boundary given by the black dashed lines for the dispersion along $k_x$ at $k_z = \pi/\Lambda$. Hence, we obtain to the band dispersions along the $k_x$ direction at $\varphi = 0°$, which linearly cross each other at $\theta = 44°$ and $f = 591\text{THz}$ for both the TE and TM polarizations. The fringe pattern on the transmission spectra intrinsically stems from the Fabry-Perot interference of the Bloch modes, for which the peak frequencies satisfy:

$$Nk_z\Lambda = m\pi \tag{4}$$

In this expression, $N = 12$ is the number of unit cells of our sample, $k_z$ is the corresponding Bloch wave vector, and $m \in \mathbb{Z}$. By the aid of Eq. (4), we are capable to obtain the dispersion along $k_z$ at $k_x = k_{\rho D}$ and $\varphi = 0°$, as shown in Figs. 2e and 2f, in which the dashed lines are theoretical results and the open circles come from the peak data (details can be found in Supplementary information Sec. IV). Figure 2 indicates that the two bands cross each other linearly along both the $k_x$ and $k_z$ directions for either the TE or the TM polarization at the same degeneracy point. Considering that the band dispersions are identical for all the azimuth angles, this four-fold degeneracy point actually extends to a ring shape, *i.e.*, DNR, in the $k_x - k_y$ plane. Consequently, we experimentally demonstrate the existence of photonic DNLS.

In addition to the DNR, our system exhibits a new type of nearly degenerated surface states. To demonstrate this, we deposit a silver film with a thickness of 25nm on the PC to confine light. Tamm-like surface states are formed between the silver film and PCs. Since we have a DNR, these Tamm-like surface states exist for both polarizations. It is worth noting that the Tamm-like surface can either be expanded by the DNR, or extended from the DNR to infinity, depending on the detail of the PC surface truncation. In our case, the PC is truncated with half of layer B on top, and with this setup, the composite system exhibits the surface states for both polarizations. (The Tamm-like surface states that extended from the DNR to infinity are shown in the Supplementary information, Sec. V.) Figures 3a and 3b exhibit the surface states for three typical values of $n_A d_A / n_B d_B$ for the TE and TM polarizations, respectively. Besides the bulk DNR, the surface states therein also possess a nodal point at $\Gamma$ ($k_x = 0, k_y = 0$) due to the TE and TM degeneracy protected by the rotational symmetry. Since the rotational-symmetry protected surface nodal point (SNP) is far away from the DNR within the spectrum and always beneath the DNR frequency, the resultant surface states are broadband and hence featured by a bowl-like dispersion. Accordingly, we name this new type of surface state double-bowl state (DBS). Moreover, owing to the fact that the TE and TM polarizations are degenerate at both the DNR and SNP, the surface states are nearly degenerate over the entire spectrum range (Supplementary information Sec. V). This feature is distinct from all previous topological surface states which are pinned by only a single type of topological degeneracy corresponding to either the nodal point or nodal line. Our system thus offers the possibility of generating surface states with arbitrary polarizations. As we know, the realization of a broadband degeneracy of the TE and TM polarized modes remains elusive in all other waveguides such as dielectric waveguides or surface plasmon waveguides, stemming from the fact that the electric and magnetic responses of optical materials are in general different. Therefore, this kind of degeneracy rooted in our system endows us with more freedom to manipulate photons, and it opens a novel avenue for exploring polarized states through the process of light-matter interaction.

Experimentally, the dispersion of the DBS can be identified from the reflection spectra. The experimental and simulation results for both polarizations are provided in Figs. 3c–3f. Compared with those for only the PCs (Supplementary information Sec. VI), the reflection spectra demonstrate a global increase due to the presence of the silver layer. Besides that, we can observe the emergence of a new reflection minimum inside the original bulk band gap (bounded by the black dashed lines). These resonance reflection deeps prove the existence of surface states unambiguously. We then numerically calculate the dispersion of the surface states, of which the results are displayed as the black solid lines in Figs. 3c–3f. The black solid lines coincide perfectly with the reflection deeps, which further confirm our argument.

## Discussion

In conclusion, we have experimentally demonstrated a new mechanism to realize a photonic type-II DNLS, which has no counterpart in existing electronic DNLSs where electron spin plays the role of polarization. The dispersion around the DNR is obtained through the angle-resolved transmission measurements. When the photonic DNLS is truncated properly and deposited with a silver film on top, the composite system exhibits broadband DBS for both the TE and TM polarizations. The DBS is identified through the deeps in the angle-resolved reflection spectra. Moreover, the DBS is preserved even if the silver film is replaced by another photonic DNLS with a different truncation (see Supplementary information Sec. VII). Our work suggests that photonic topological systems cannot be adequately classified by spinless space groups. On the application side and considering the extreme field concentration due to the surface states, our system can be regarded as an ideal platform for investigating phenomena that require large field enhancement, such as cavity polaritons and nonlinear optics. Additionally, since the DBS for the TE and TM polarizations are almost degenerate over a large spectrum range, this platform exhibits unique advantages in investigating the light-matter interaction between circular polarized photons and spin or valley electrons in a condensed matter system, such as spin polaritons in a microcavity[37] or valley electrons in $MoS_2$[38].

## Materials and methods

### Experiments

**A. Sample fabrication**

In our experiments, two sets of PC samples were fabricated with electron beam evaporation on top of a 1.0mm thick $SiO_2$ substrate. The first sample was used to measure the bulk band dispersion, and the second sample to validate the surface state. For the first sample used, as shown in Fig. 2, 12 unit cells with alternating layers of $SiO_2$ (layer A, $d_A = 388\text{nm}$) and $Ta_2O_5$ (layer B, $d_B = 597\text{nm}$) were deposited, with a full layer B at the bottom. The scanning electron microscope image of this sample is shown in Fig. 1a, for which the uncertainty of the thickness is under 10nm. For the second sample used, as shown in Fig. 3, 12 unit cells were first deposited on the $SiO_2$ substrate with a full layer B as the first layer. Then an additional layer B with half the thickness $d_B/2$ was deposited on the top in order to control the dispersion of the surface states. The geometric parameters of the second PC were $d_A = 402\text{nm}$ and $d_B = 605\text{nm}$. After that, a silver film with a thickness of $(25\pm5)$ nm was deposited on top of the PC sample with electron beam evaporation.

**B. Spectra measurement**

The transmission and reflection spectra were measured at room temperature with an Ideaoptics Instrument PG2000-Pro spectrometer (370–1050nm) with a wavelength resolution of 0.35nm. The polarization of the incident wave and the transmission and reflection waves were selected with a polarizer. The measurements were performed as the incident angle varies from $0^o$ to $60^o$ at intervals of $0.5^o$. To increase the accuracy and the stability of our measurements, we set the integration time to 200ms, and we averaged over five independent measurements for both polarizations.

**Simulations**

All the simulations for the transmission and reflection spectra were performed with Lumerical FDTD Solutions, a commercial software based on the Finite-Different Time-Domain Method. A two-dimensional geometry was exploited for which the *z*-axis was chosen as the stacking direction and the *x*-axis represented an arbitrary direction parallel to the surface of layers. Air serves as the background medium. The refractive indexes of $SiO_2$ and $Ta_2O_5$ were extracted from the measured data (see Supplementary Data I). The relative permittivity of Ag was from the tabulated reference[39]. Fourier periodic boundary conditions were applied in the *x*-direction and perfectly matched layer conditions were introduced for the *z* termini. A plane-wave source was placed at the top boundary, and it generated polarized waves within the frequency regime of interest (530–630THz). To achieve angle-resolved transmission and reflection spectra, we swept the incident angle from $0^o$ to $60^o$ at intervals of $0.5^o$ for both the TE and TM polarizations, and used two frequency-domain field and power monitors located far away from the structure to record the spectra. All the simulations were performed at standard temperature and pressure (STP).

## Figures

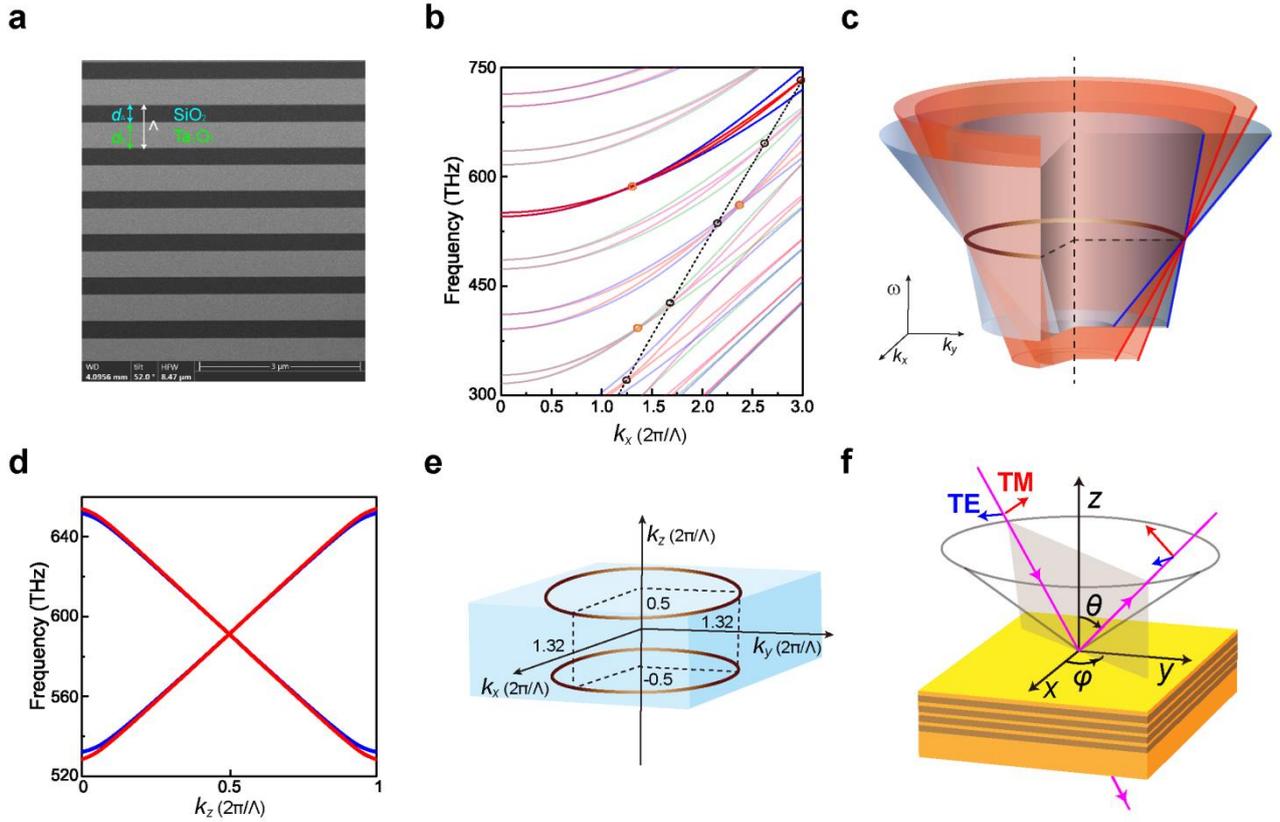

**Fig. 1 Photonic DNLS. a** SEM image of the PC. **b** Dispersion along the $k_x$ direction at $k_z = 0$ (green and magenta) and $k_z = \pi/\Lambda$ (red and blue). The blue and magenta lines represent TE polarization, and the red and green lines represent the TM polarization. The bands forming the DNR, which is the focus in our study, are highlighted, while the other bands are partially transparent. The dashed line indicates the locus of the Brewster angles, where the TM gaps close. The gap closing points for both polarizations (at the Brewster angles) are encircled in orange (black). **c** Sketch of the in-plane dispersion around the DNR (golden). The DNR is four-fold degenerate, consisting of two sets of type-II Weyl nodal rings of TE (blue) and TM (red) polarizations. **d** Dispersions along the $k_z$ direction around the four-fold degeneracy point for TE (blue) and TM (red) polarizations. **e** The position of the DNR in the reciprocal space. **f** The experimental set-up. We employ SiO$_2$ (Ta$_2$O$_5$) for layer A (B), with a thickness of $d_A$=388nm ($d_B$=597nm). The refractive index of the SiO$_2$ (Ta$_2$O$_5$) is around 1.48 (2.25) with light dispersion in the visible regime (see Supplementary Data I).

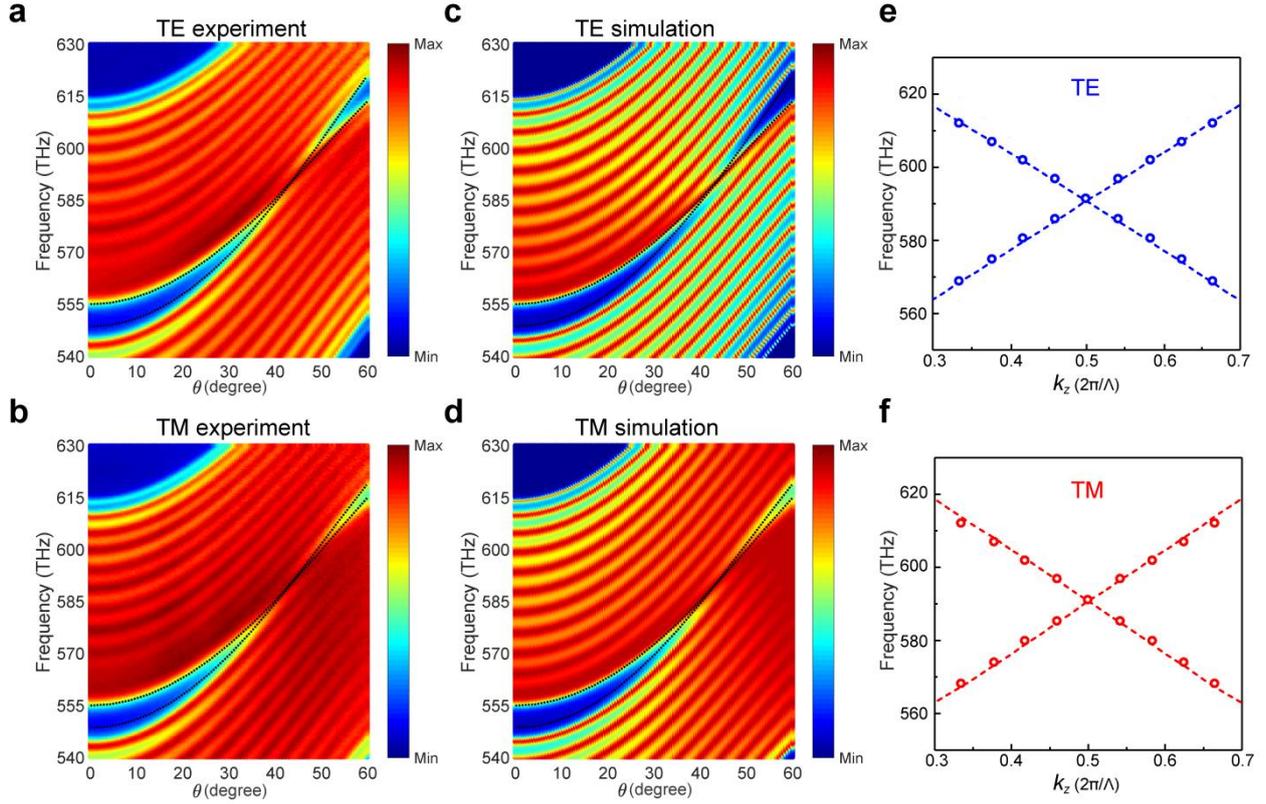

**Fig. 2 Experimental observation of photonic DNR. a, b** Measured and **c, d** simulated transmission spectra for the PC sample as the incident angle varied from $0^o$ to $60^o$ for the TE and TM plane-wave excitations. The transmission spectra fill the region bounded by the bulk bands (black dashed lines) along the $k_x$ direction at $k_z = \pi/\Lambda$. **e, f** Extracted dispersions along $k_z$ from the Fabry-Perot interference pattern for **e** TE and **f** TM polarizations, in which the open circles are acquired from the transmission peaks. The dashed lines are theoretical results for reference. The dielectric and geometric parameters of the PC are the same as those shown in Fig. 1.

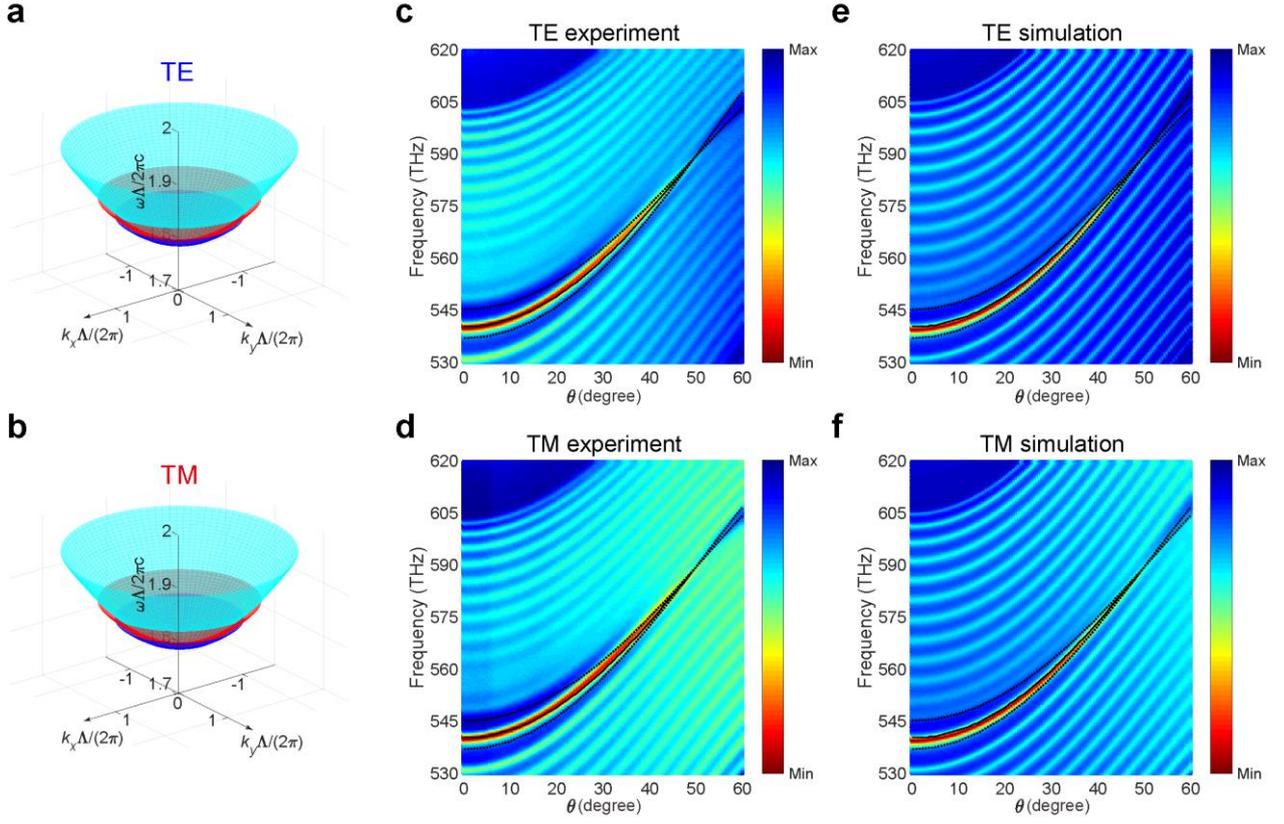

**Fig. 3 Observation of the DBS. a, b** Bowl surface state dispersion for **a** TE and **b** TM polarizations, in which the cyan, red, and dark blue surfaces indicate cases in which $n_A d_A / n_B d_B$ equals 0.442, 0.422, and 0.415, respectively. $n_A d_A + n_B d_B$ remains constant in these simulations. In the experiment, $n_A d_A / n_B d_B$ of our PC is 0.442. **c, d** Measured and **e, f** simulated reflection spectra for the silver film/PC sample as the incident angle changes from $0^o$ to $60^o$ for TE and TM excitations. The black solid lines mark the numerically simulated bowl surface states dispersion, while the black dashed lines show the dispersion of the bulk PC bands. The PC shares identical dielectric properties with those shown in Fig. 2 but with slightly different thicknesses of $d_A$=402nm and $d_B$=605nm. The top layer of the truncated PC is layer B with $d_B/2$. The thickness of the silver film on the top is 25nm ($\pm 5$nm).

## Acknowledgments

This work is supported by National Natural Science Foundation of China (Grant No. 11904264, 11690033), National Key Research and Development Program of China (No. 2017YFA0205700) and National Key R&D Program of China (2017YFA0303702). M. X. is supported by the startup funding of Wuhan University.

## Author contributions

M. H. did the numerical simulations. M. H., Y. Z., X. J., and T. Q. carried out the experiments. M. X., S.N.Z and H. L. supervised the project. M. H., M. X. and H. L. wrote the manuscript. All the authors contributed to the analysis and discussion of the results.

## Conflict of interest

The authors declare that they have no conflict of interest.

# References

1. Qi, X. L. & Zhang, S. C. Topological insulators and superconductors. *Reviews of Modern Physics* **83**, 1057-1110 (2011).

2. Hasan, M. Z. & Kane, C. L. *Colloquium*: topological insulators. *Reviews of Modern Physics* **82**, 3045-3067 (2010).

3. Lu, L. *et al.* Weyl points and line nodes in gyroid photonic crystals. *Nature Photonics* **7**, 294-299 (2013).

4. Lu, L. *et al.* Experimental observation of Weyl points. *Science* **349**, 622-624 (2015).

5. Ozawa, T. *et al.* Topological photonics. *Reviews of Modern Physics* **91**, 015006 (2019).

6. Ma, G. C., Xiao, M. & Chan, C. T. Topological phases in acoustic and mechanical systems. *Nature Reviews Physics* **1**, 281-294 (2019).

7. Chiu, C. K. *et al.* Classification of topological quantum matter with symmetries. *Reviews of Modern Physics* **88**, 035005 (2016).

8. Bradlyn, B. *et al.* Topological quantum chemistry. *Nature* **547**, 298-305 (2017).

9. Tang, F. *et al.* Comprehensive search for topological materials using symmetry indicators. *Nature* **566**, 486-489 (2019).

10. Vergniory, M. G. *et al.* A complete catalogue of high-quality topological materials. *Nature* **566**, 480-485 (2019).

11. Zhang, T. T. *et al.* Catalogue of topological electronic materials. *Nature* **566**, 475-479 (2019).

12. Watanabe, H. & Lu, L. Space group theory of photonic bands. *Physical Review Letters* **121**, 263903 (2018).

13. Xiong, Z. F. *et al.* Hidden-symmetry-enforced nexus points of nodal lines in layer-stacked dielectric photonic crystals. *Light: Science & Applications* **9**, 176 (2020).

14. Fang, C. *et al.* Topological nodal line semimetals with and without spin-orbital coupling. *Physical Review B* **92**, 081201(R) (2015).

15. Carter, J. M. *et al.* Semimetal and topological insulator in perovskite Iridates. *Physical Review B* **85**, 115105 (2012).

16. Kim, Y. *et al.* Dirac line nodes in inversion-symmetric crystals. *Physical Review Letters* **115**, 036806 (2015).

17  Weng, H. M. *et al*. Topological node-line semimetal in three-dimensional graphene networks. *Physical Review B* **92**, 045108 (2015).

18  Mullen, K., Uchoa, B. & Glatzhofer, D. T. Line of Dirac nodes in hyperhoneycomb lattices. *Physical Review Letters* **115**, 026403 (2015).

19  Liu, Z. K. *et al*. Discovery of a three-dimensional topological Dirac semimetal, $Na_3Bi$. *Science* **343**, 864-867 (2014).

20  Xu, S. Y. *et al*. Discovery of a Weyl fermion semimetal and topological Fermi arcs. *Science* **349**, 613-617 (2015).

21  Burkov, A. A., Hook, M. D. & Balents, L. Topological nodal semimetals. *Physical Review B* **84**, 235126 (2011).

22  Xiao, M. *et al*. Experimental demonstration of acoustic semimetal with topologically charged nodal surface. *Science Advances* **6**, eaav2360 (2020).

23  Belopolski, I. *et al*. Discovery of topological Weyl fermion lines and drumhead surface states in a room temperature magnet. *Science* **365**, 1278-1281 (2019).

24  Yan, Q. H. *et al*. Experimental discovery of nodal chains. *Nature Physics* **14**, 461-464 (2018).

25  Zhang, A. M. *et al*. Interplay of Dirac electrons and magnetism in $CaMnBi_2$ and $SrMnBi_2$. *Nature Communications* **7**, 13833 (2016).

26  Rhim, J. W. & Kim, Y. B. Landau level quantization and almost flat modes in three-dimensional semimetals with nodal ring spectra. *Physical Review B* **92**, 045126 (2015).

27  Huh, Y., Moon, E. G. & Kim, Y. B. Long-range Coulomb interaction in nodal-ring semimetals. *Physical Review B* **93**, 035138 (2016).

28  Shao, Y. M. *et al*. Electronic correlations in nodal-line semimetals. *Nature Physics* **16**, 636-641 (2020).

29  Guo, Q. H. *et al*. Observation of three-dimensional photonic Dirac points and spin-polarized surface arcs. *Physical Review Letters* **122**, 203903 (2019).

30  Cai, X. X. *et al*. Symmetry-enforced three-dimensional Dirac phononic crystals. *Light: Science & Applications* **9**, 38 (2020).

31  Cheng, H. B. *et al*. Discovering topological surface states of Dirac points. *Physical Review Letters* **124**, 104301 (2020).

32  Deng, W. Y. *et al.* Nodal rings and drumhead surface states in phononic crystals. *Nature Communications* **10**, 1769 (2019).

33  Qiu, H. H. *et al.* Straight nodal lines and waterslide surface states observed in acoustic metacrystals. *Physical Review B* **100**, 041303(R) (2019).

34  Yang, E. C. *et al.* Observation of non-abelian nodal links in photonics. *Physical Review Letters* **125**, 033901 (2020).

35  Xiao, M., Zhang, Z. Q. & Chan, C. T. Surface impedance and bulk band geometric phases in one-dimensional systems. *Physical Review X* **4**, 021017 (2014).

36  Soluyanov, A. A. *et al.* Type-II Weyl semimetals. *Nature* **527**, 495-498 (2015).

37  Kavokin, K. V. *et al.* Quantum theory of spin dynamics of exciton-polaritons in microcavities. *Physical Review Letters* **92**, 017401 (2004).

38  Zeng, H. L. *et al.* Valley polarization in $MoS_2$ monolayers by optical pumping. *Nature Nanotechnology* **7**, 490-493 (2012).

39  Palik, E. D. Handbook of Optical Constants of Solids. (Orlando: Academic Press, 1985).

# Supplementary Information for:
# Double-bowl State in photonic Dirac nodal line semimetal

Mengying Hu[1], Ye Zhang[1], Xi Jiang[1], Tong Qiao[1], Qiang Wang[2], Shining Zhu[1], Meng Xiao[3], and Hui Liu[1]

[1]National Laboratory of Solid State Microstructures, School of Physics, Collaborative Innovation Center of Advanced Microstructures, Nanjing University, Nanjing 210093, China.

[2]Division of Physics and Applied Physics, School of Physical and Mathematical Sciences, Nanyang Technological University, Singapore 637371, Singapore.

[3]Key Laboratory of Artificial Micro- and Nano-structures of Ministry of Education and School of Physics and Technology, Wuhan University, Wuhan 430072, China.

Correspondence: Meng Xiao (phmxiao@whu.edu.cn); Hui Liu (liuhui@nju.edu.cn).

## Section I: Bands crossing condition for off-normal directions

For the normal direction ($k_x = k_y = 0$), the band structure is given by

$$\cos(k_z\Lambda) = \cos(k_0 n_A d_A)\cos(k_0 n_B d_B) - \frac{1}{2}\left(\frac{\zeta_A}{\zeta_B} + \frac{\zeta_B}{\zeta_A}\right)\sin(k_0 n_A d_A)\sin(k_0 n_B d_B) \quad \text{(S1)}$$

where $k_z$ is the Bloch wave vector, $\Lambda$ is the unit cell length, $k_0$ is the wave vector in vacuum, $\zeta_i = \sqrt{\mu_i/\varepsilon_i}$, $n_i$ and $d_i$ denote respectively, the impedance, refractive index and the thickness of layer $i$ ($i$ = A or B). For normal incidence, TE and TM modes exhibit the same band structure. The accidental degeneracies occur at $n_A d_A / n_B d_B = m_1/m_2 \in \mathbb{Q}$, where $\{m_1, m_2\} \in \mathbb{N}^+$. As a result, the $(m_1 + m_2)^{\text{th}}$ band and the $(m_1 + m_2 + 1)^{\text{th}}$ band cross at $f_{m_1+m_2} = (m_1 + m_2)c/2(n_A d_A + n_B d_B)$, where $c$ is the speed of light in vacuum. Note here, the band crossing condition is independent of the impedance and only depends on the ratio of optical path inside each layer. This bands crossing condition for the normal direction have been analytically derived in Ref. [1].

In this section, we show that this kind of crossing condition can be extended to off-normal directions. Since the system is isotropic along in-plane directions, we set $k_y = 0$ for simplicity when we consider off-normal directions. The band structure of a dielectric binary PC with a certain $k_x$ can be written as:

$$\cos(k_z\Lambda) = \cos(k_{zA} d_A)\cos(k_{zB} d_B) - \frac{1}{2}\left(\frac{\mu_B k_{zA}}{\mu_A k_{zB}} + \frac{\mu_A k_{zB}}{\mu_B k_{zA}}\right)\sin(k_{zA} d_A)\sin(k_{zB} d_B) \quad \text{(S2)}$$

for TE polarized mode, and

$$\cos(k_z\Lambda) = \cos(k_{zA} d_A)\cos(k_{zB} d_B) - \frac{1}{2}\left(\frac{\varepsilon_B k_{zA}}{\varepsilon_A k_{zB}} + \frac{\varepsilon_A k_{zB}}{\varepsilon_B k_{zA}}\right)\sin(k_{zA} d_A)\sin(k_{zB} d_B) \quad \text{(S3)}$$

for TM polarized mode. Here, $k_{zi} = \sqrt{k_i^2 - k_x^2}$ ($i$ = A or B). It is noteworthy that only $k_{zi}$ in Eqs. (S1) and (S2) depends on $k_x$. For normal incidence, $k_{zi}$ is proportional to $n_i$, i.e., $k_{zi} = n_i k_0$. As for the off-normal directions, we define an effective refractive index which describes the propagation phase delay along the $z$ direction as

$$\tilde{n}_i = \sqrt{\varepsilon_i \mu_i - k_x^2/k_0^2} \quad \text{(S4)}$$

such that $k_{zi} = \sqrt{k_i^2 - k_x^2} = \tilde{n}_i k_0$. Comparing Eqs. (S2) and (S3) with Eq. (S1), we can see that they all exhibit the same form if we redefine $\tilde{\zeta}_i^{\text{TE}} \equiv \mu_i / \tilde{n}_i$, $\tilde{\zeta}_i^{\text{TM}} \equiv \tilde{n}_i / \varepsilon_i$ ($i = $ A or B). Thus the bands crossing condition for off-normal directions in both Eqs. (S2) and (S3) can be obtained by replacing $n_i$ with $\tilde{n}_i$, i.e.,

$$\tilde{n}_A d_A / \tilde{n}_B d_B = m_1 / m_2 \in \mathbb{Q}, \tag{S5}$$

Meanwhile, the $(m_1 + m_2)^{\text{th}}$ band and the $(m_1 + m_2 + 1)^{\text{th}}$ band cross at $f_{m_1+m_2} = (m_1 + m_2)c / 2(\tilde{n}_A d_A + \tilde{n}_B d_B)$. For the TM modes, the bands also cross at the Brewster angle when $\tilde{\zeta}_i^{\text{TM}} \equiv \mu_i / \tilde{n}_i = 1$.

Above we show that the four-fold degeneracy condition is given in Eq. (S5), and here we proceed to show that Eq. (S5) is robust against material dispersions as well as the variation of geometric parameters. Firstly, one can find that rhs of Eq. (S5) is a constant, and lhs of Eq. (S5) is a continuous function of $k_x$ and system parameters. Note here, $k_0$ of the degenerate point is related to $k_x$ through $k_x^2 + k_z^2 = k_0^2$. Meanwhile, Eq. (S5) is satisfied at isolated $k_x$s, thus the only possibility is that the lhs as a continuous function of $k_x$ crosses rhs, a constant once at the band degeneracy. In the presence of a small change of the system parameters such as dispersions or thickness, Equation (S5) can still be satisfied with a shift of $k_x$. As an example, we consider the case where the thickness of two layers change as $d'_A = (1+\alpha)d_A$ and $d'_B = (1+\alpha)d_B$ with $\alpha$ being a small number. Here $d_A = 388$nm and $d_B = 597$nm are the thickness of the SiO$_2$ and Ta$_2$O$_5$ layers, respectively. The presence of $\alpha$ describes a common system error in fabricating layered structures. Same as the main text, we focus on the four-fold degeneracy point at 591THz. The locations of the four-fold degeneracy point as a function of $\alpha$ is shown in Fig. S1. From Fig. S1, we can see that the four-fold degeneracy point preserves and shifts in $k_x$ as we vary $\alpha$. SiO$_2$ and Ta$_2$O$_5$ are almost dispersionless within the frequencies of interest. (see Supplementary Data I) Nevertheless, we want to mention that the above argument also works for dispersive material. The four-fold degeneracies preserve and the dispersion of material only introduce a small shift of $k_x$, similar as the effect of $\alpha$ discussed above.

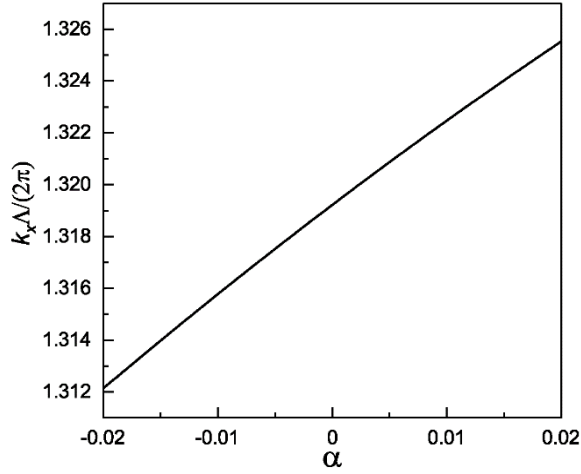

Fig. S1 $k_x$ of the four-fold degeneracy point as a function of $\alpha$.

Since this system exhibits time reversal symmetry and is in-plane isotropic, the band dispersion should be symmetric for $k_z$ with respect to $k_z = 0$ at an arbitrary fixed $k_x$. Then we can see that the band dispersions should be monotonic as functions of $k_z$ for $k_z > 0$ or $k_z < 0$ at that fixed $k_x$, otherwise there will be at least four $k_z$s (two for $k_z > 0$ and two for $k_z < 0$) for one frequency in the non-monotonic region of the band dispersion. This is contradictory to Eqs. (S2) and (S3) since for each frequency, the rhs of Eqs. (S2) or (S3) is single valued and hence we have at most two $k_z$s (one positive and one negative) on the lhs. Thus the monotonicity along $k_z$ further leads to the conclusion that the degenerate points of two bands occur only at the zone boundary ($k_z = \pi/\Lambda$) or the zone center ($k_z = 0$).

**Section II: Existence of Dirac nodal ring at $k_z = 0$**

In the main text, we analyze a Dirac nodal ring (DNR) at $k_z = \pi/\Lambda$ whose location is shown with the golden ring in Fig. 1e. Here in this section, we provide the band dispersion for another DNR at $k_z = 0$. Here the parameters we use are the same as Fig. 1 in the main text, and the DNR corresponds to the orange circle at a lower frequency (around 400THz) in Fig. 1b. Figs. S2a and S2b show the band dispersion along the $k_x$ and $k_z$ direction around the four-fold degeneracy point. Considering the fact that the system are rotational invariant for in-plane directions, we thus get a DNR located at $k_z = 0$ as sketched in Fig. S2c.

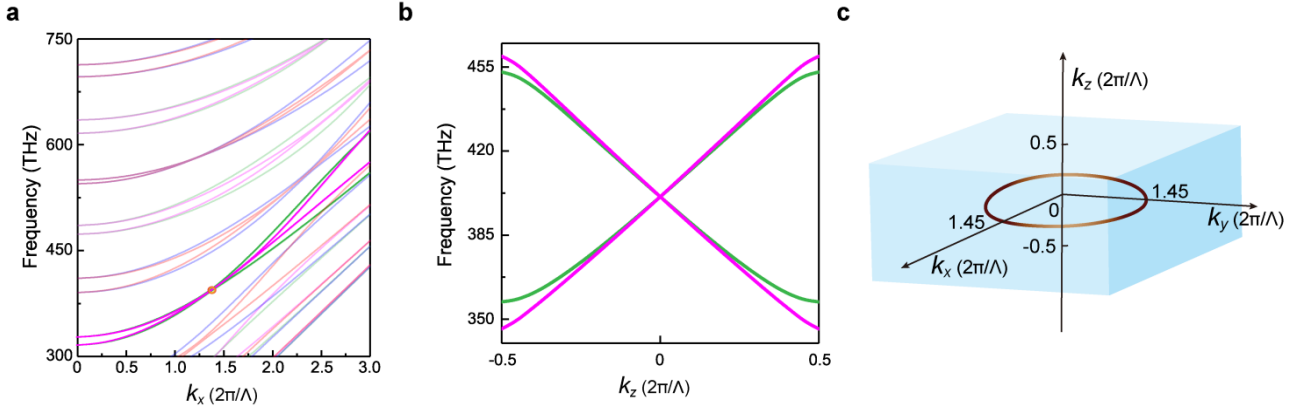

Fig. S2 Band dispersion along **a** $k_x$ and **b** $k_z$ for a DNR at $k_z = 0$ (green and magenta). The magenta and blue (olive and red) lines represent the band dispersion for TE (TM) polarization. **c**. Sketch of the Dirac nodal ring in the momentum space.

## Section III: Transmission spectra at different $\varphi$ s

To verify the in-plane isotropy of our system, we measured transmission spectra at various azimuthal angles $\varphi$s besides $\varphi = 0°$ shown in the main text. Fig. S3 displays the transmission spectra for both TE and TM polarizations at $\varphi = 0°$, $\varphi = 45°$ and $\varphi = 90°$, which are almost the same, sufficiently evidencing the isotropy inherent in our system.

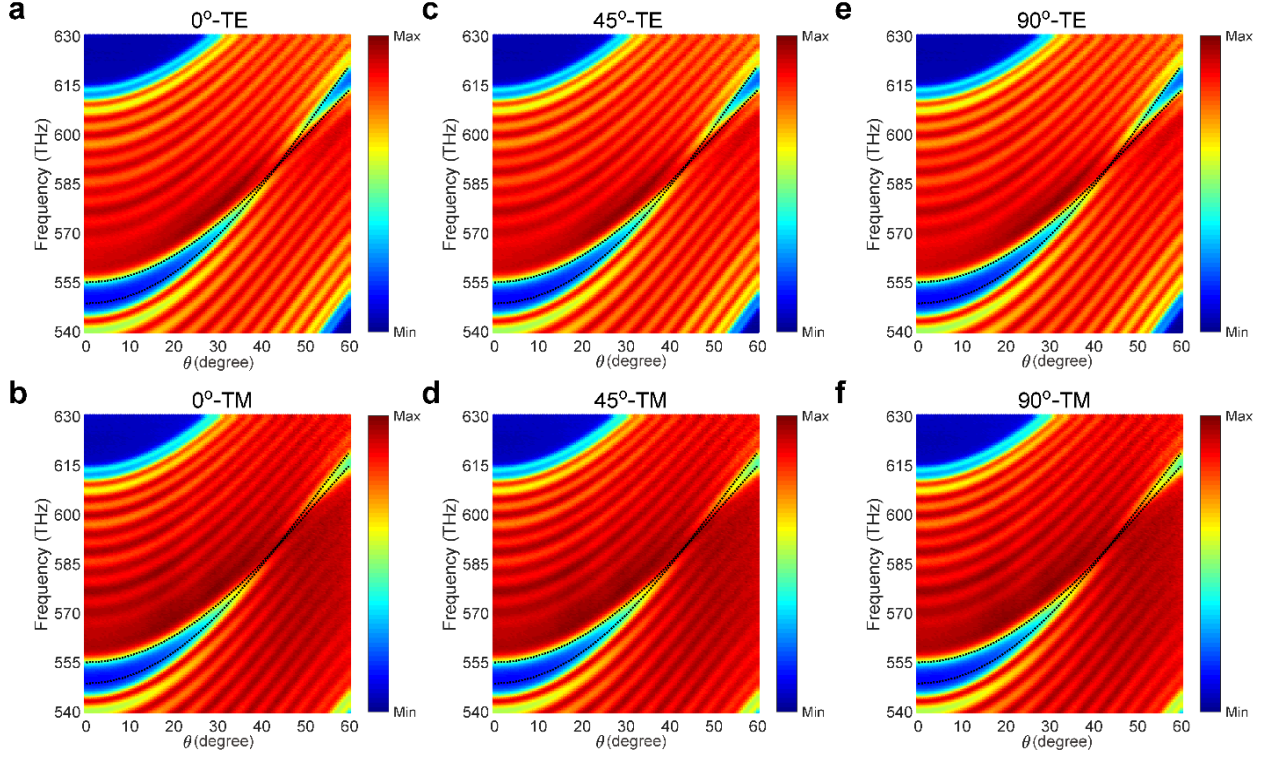

Fig. S3 Measured transmission spectra for both polarizations as **a, b**, $\varphi = 0°$ **c, d** $\varphi = 45°$ and **e, f** $\varphi = 90°$. Black dashed lines mark the band dispersion at $k_z = \pi/\Lambda$.

## Section IV: Band dispersion extracted from the Fabry-Perot interference

In this section, we offer details about the approach to extract the band dispersion at the DNR along the $k_z$ direction as shown in Fig. 2. To obtain the dispersion along the $k_z$ direction at the DNR, we need the transmission and reflection spectra at fixed $k_{\rho D} = 1.32(2\pi/\Lambda)$. However, the experimental data are measured at fixed incident angle, thus we first smoothly interpolate the measured spectrum, and then extract the transmission and reflection spectra at fixed $k_{\rho D}$, and the data are shown in Fig. S4a.

The geometry of our system is shown in Fig. S4b, with the number of unit cell $N$=12. The Fabry-Perot interference pattern for a fixed $k_z$ actually comes from the interference of multiple Bloch modes. Following a similar transfer matrix method used in Ref. [2], we obtain the reflection and transmission coefficients of the system as

$$R = \left| \frac{(\chi_1 - 1)\sin[(N-1)k_z\Lambda] - (\chi_1(\chi_2 + \chi_3) - \chi_2 + \chi_3)\sin(Nk_z\Lambda)}{(\chi_1 + 1)\sin[(N-1)k_z\Lambda] - (\chi_1(\chi_2 + \chi_3) + \chi_2 - \chi_3)\sin(Nk_z\Lambda)} \right|^2 \tag{S6}$$

and

$$T = 1 - R, \tag{S7}$$

with

$$\begin{cases} \chi_2^{\mathrm{TE}} = \dfrac{k_{z0}}{k_{zA}} \\ \chi_2^{\mathrm{TE}} = e^{ik_{zA}d_A}\left( \cos(k_{zB}d_B) + \dfrac{i}{2}\left( \dfrac{\tilde{\zeta}_A^{\mathrm{TE}}}{\tilde{\zeta}_B^{\mathrm{TE}}} + \dfrac{\tilde{\zeta}_B^{\mathrm{TE}}}{\tilde{\zeta}_A^{\mathrm{TE}}} \right)\sin(k_{zB}d_B) \right) \\ \chi_3^{\mathrm{TE}} = e^{ik_{zA}d_A}\left( -\dfrac{i}{2}\left( \dfrac{\tilde{\zeta}_A^{\mathrm{TE}}}{\tilde{\zeta}_B^{\mathrm{TE}}} - \dfrac{\tilde{\zeta}_B^{\mathrm{TE}}}{\tilde{\zeta}_A^{\mathrm{TE}}} \right)\sin(k_{zB}d_B) \right) \end{cases} \tag{S8}$$

and

$$\begin{cases} \chi_2^{\mathrm{TM}} = \dfrac{n_A^2 k_{z0}}{n_0^2 k_{zA}} \\ \chi_2^{\mathrm{TM}} = e^{ik_{zA}d_A}\left( \cos(k_{zB}d_B) + \dfrac{i}{2}\left( \dfrac{\tilde{\zeta}_A^{\mathrm{TM}}}{\tilde{\zeta}_B^{\mathrm{TM}}} + \dfrac{\tilde{\zeta}_B^{\mathrm{TM}}}{\tilde{\zeta}_A^{\mathrm{TM}}} \right)\sin(k_{zB}d_B) \right) \\ \chi_3^{\mathrm{TM}} = e^{ik_{zA}d_A}\left( -\dfrac{i}{2}\left( \dfrac{\tilde{\zeta}_A^{\mathrm{TM}}}{\tilde{\zeta}_B^{\mathrm{TM}}} - \dfrac{\tilde{\zeta}_B^{\mathrm{TM}}}{\tilde{\zeta}_A^{\mathrm{TM}}} \right)\sin(k_{zB}d_B) \right) \end{cases} \tag{S9}$$

for TE and TM polarizations, respectively. Here $k_{z0}$ is the z-component of the wave vector in the air. The comparison between these analytical results (solid lines) and numerical simulations (open circles) are provided in Fig. S4c. In Eq. S7 (S6), the transmission peak (reflection deep) is reached when

$$\sin(Nk_z\Lambda) = 0 \tag{S10}$$

or

$$Nk_z\Lambda = m\pi, \tag{S11}$$

where $m$ is an integer. Under such a condition, the transmission maximum is simplified as

$$T_{\max} = \frac{4\chi_1}{(1+\chi_1)^2} \tag{S12}$$

$T_{max}$ as functions of frequency are for both polarization also provided as dashed lines in Fig. S4c for comparison. Equation (S11) builds a relation between the values of Bloch wave vector $k_z$ and the corresponding frequencies at which $T$ ($R$) reaches the peak (deep) value, and which then gives the band dispersion we extract and provided in Figs. 2e and 2f in the main text.

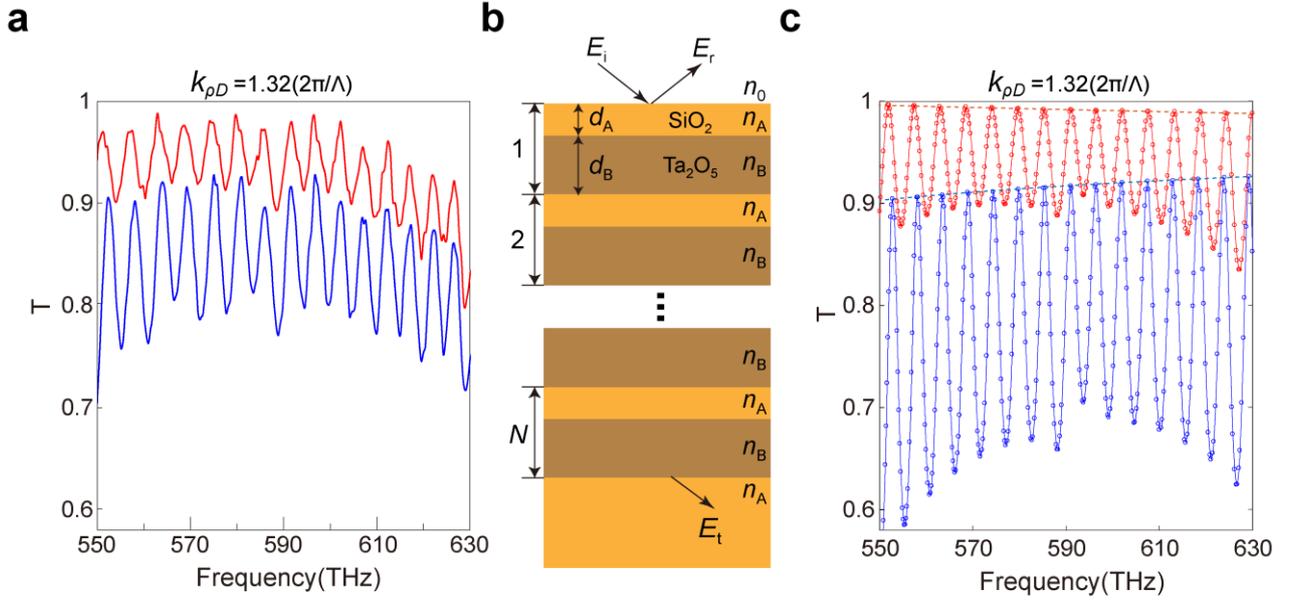

Fig. S4 **a.** Measured transmission spectra for TE (blue) and TM (red) polarizations at $k_{\rho D} = 1.32(2\pi/\Lambda)$. **b.** Sketch of our system. The plane wave incidents on the sample from air ($n_0 = 1$) and transmit into the substrate made of SiO$_2$. The incident, reflected and transmitted amplitudes of electric field are labeled as $E_i$, $E_r$ and $E_t$, respectively. **c.** Analytical (solid lines) and numerical (open circles) results of transmission spectra at $k_{\rho D} = 1.32(2\pi/\Lambda)$, together with the functions $T_{max}$ (dashed lines) for TE (blue) and TM (red) polarizations.

### Section V: Nearly degeneracy of the double-bowl surface states

A double-bowl surface state consists of two drumhead surface states for TE and TM polarizations, and these two drumhead surface states are degenerate at $\Gamma$ and DNR. Actually, these two drumhead surface states are almost degenerate over the whole spectra range as revealed in Fig. S5, where we render the trajectories in $k_x - f$ plane for three sets of drumhead surface states corresponding to those in Figs. 3a and 3b. Meanwhile, such a nearly ideal degeneracy can be extended all over the spectrum by tuning $n_A d_A / n_B d_B$, which is experimentally feasible. Therefore, our scheme holds great potential for generating drumhead surface states with arbitrary polarizations.

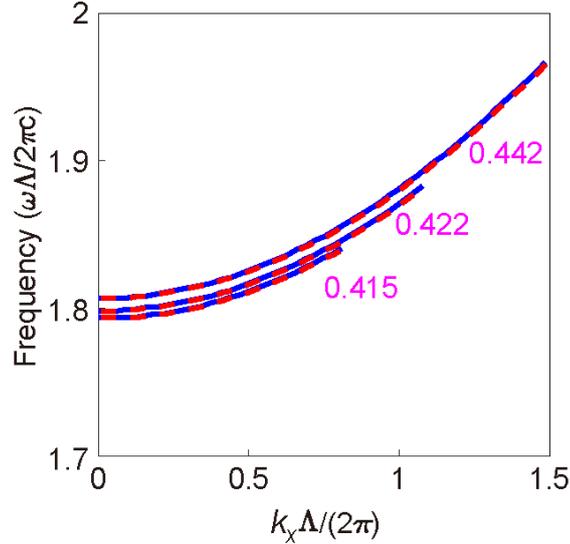

Fig. S5 Drumhead surface states with TE (blue solid lines) and TM (red dashed lines) polarizations as $n_A d_A / n_B d_B$ from top to bottom equals 0.442, 0.422 and 0.415, respectively. The sum $n_A d_A + n_B d_B$ remains constant in these three cases.

Moreover, the surface states can either be expanded by the DNR or extended from the DNR to infinity depending on the detail of the PC surface truncation. For the case shown in the main text, the PC is truncated with layer B of thickness $d_B/2$ on top, wherein the composite system exhibits drumhead surface states for both TE and TM polarizations. Here, we fabricate another PC sample truncated with a complete layer A (other parameters are identical to the sample used in Fig. 2), and on top of which we deposit a 25nm silver film. With the aid of angle-resolved transmission measurements, we are capable to achieve surface states extended from the DNR to infinity, which display themselves as transmission peaks inside the original bulk band gap (bounded by the black dashed lines) in Fig. S6.

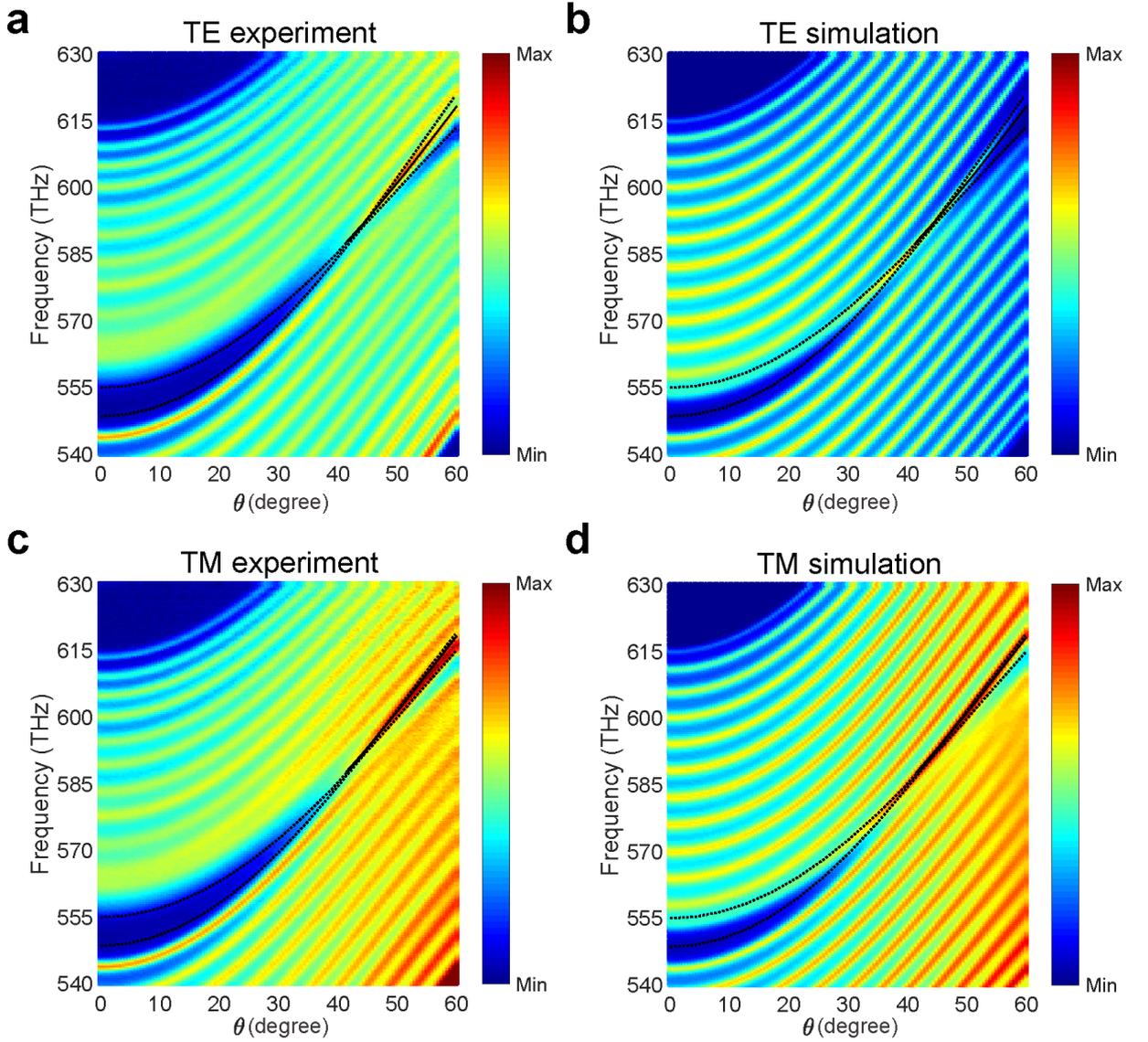

Fig. S6 Surface states under oblique-incident excitations. **a, c**, measured and **b, d**, simulated transmission spectra for **a, b**, TE and **c, d**, TM excitations, in which the surface states extend from the DNR to infinity. Black solid lines exhibit eigenfrequencies of the surface states, while black dashed lines mark the boundary of the projected band regions.

### Section VI: Reflection spectra of the PC with/without a silver film

Here in this section we provide the reflection spectra of the PC before deposited with a silver film. The parameters of the PC are the same as that used in Fig. 3. Here Figs. S7a and S7b (Figs. S7c and S7d) show the reflection spectra without (with) the silver film for the TE and TM polarization, respectively. It can be seen that, when deposited with a silver film, there is a global increase of the reflection spectra. Note here for clearance, we show the absolute values of the reflectance. Meanwhile, there is an additional reflection deep emerging inside the original band gap which corresponds to the double-bowl states.

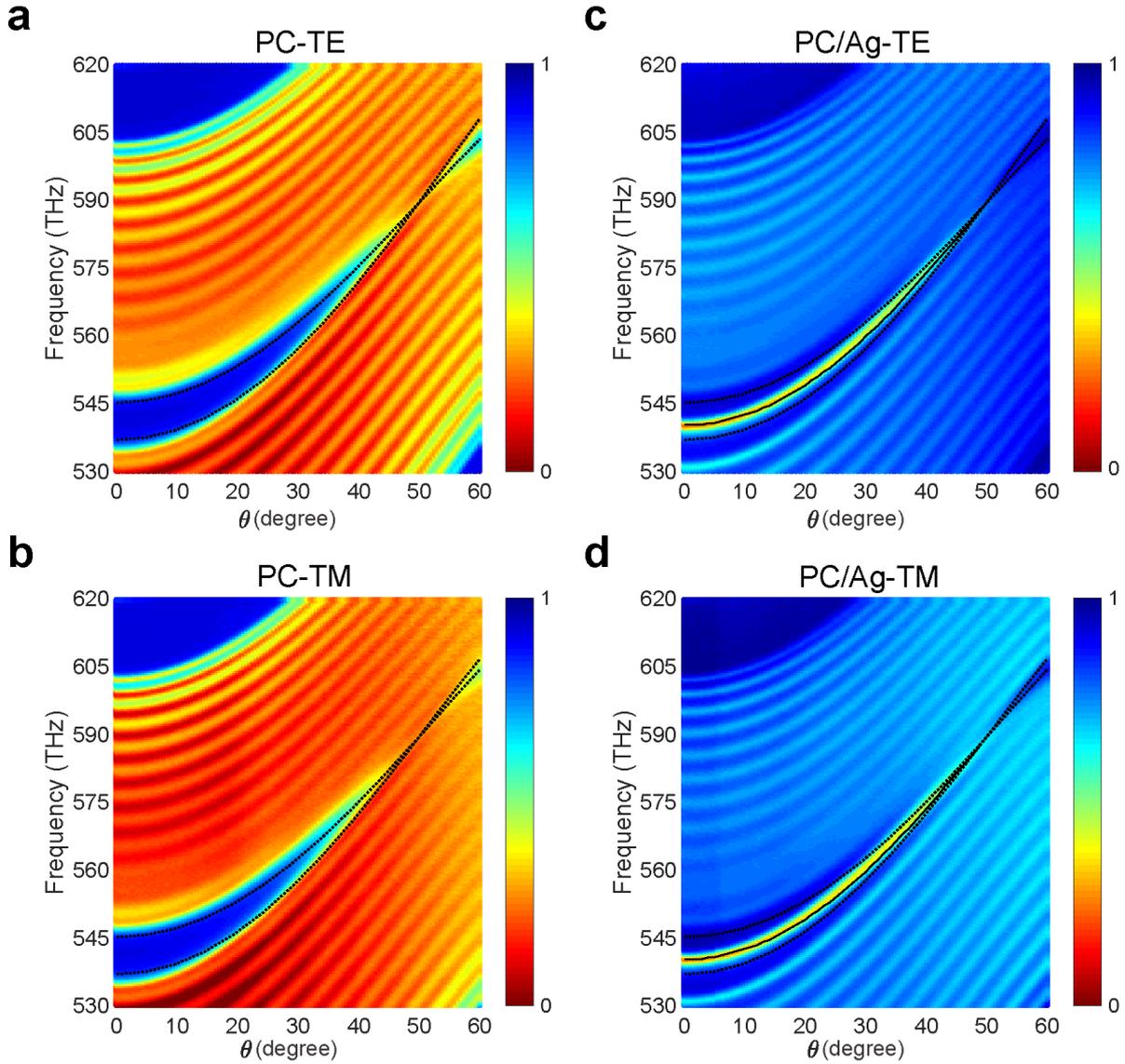

Fig. S7 Reflection spectra for the PC without (**a**, **b**) and with (**c**, **d**) a silver film of 25nm deposited on top. The PC studied here is the same as the PC in Fig. 3 of the main text. Black dashed lines correspond to the boundaries of projected bands, and the black solid line in **c** and **d** mark the dispersion of the double-bowl states.

**Section VII: Double-bowl surface states between two photonic DNLSs**

In this section we show that the double-bowl surface states can also be supported by interface between two photonic DNLSs. Here these two photonic DNLSs are the same only with different truncations at the interface. As shown in Fig. S8a, the silver film is replaced with another photonic DNLS whose first layer is SiO$_2$ with half the thickness $d_A/2$. Besides that, we also assume the two photonic DNLS are semi-infinite. The projected band together with the double-bowl surface states are shown in Fig. S8b, where the blue (red) region depicts the projected band of TE (TM) polarization, and the lines represent the surface states localized between two photonic DNLS. It is intriguing to see that such a configuration exhibits two sets of surface states: one is the double-bowl surface states and the other surface states extend from the DNR to infinity.

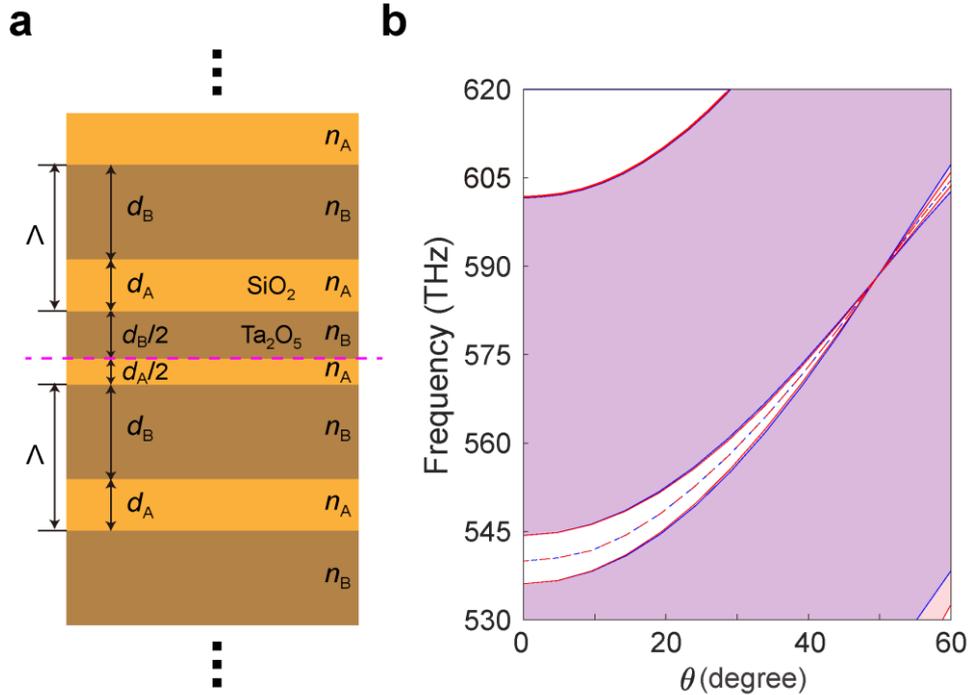

Fig. S8 **a**. Interface formed between two same photonic DNLS with different truncations. **b**. Projected band structure and the surface state dispersion. Here the red and blue regions represent the projected band of TE and TM polarizations, respectively, and the blue (red) dashed line is the dispersion of the surface states with TE (TM) polarization. Here, the thickness of the SiO$_2$ layer and Ta$_2$O$_5$ layer are $d_A = 402$nm and $d_B = 605$nm, respectively, and the refractive indexes are provided in Supplementary Data I.

## Supplementary Data I: Refractive indexes of $SiO_2$ and $Ta_2O_5$

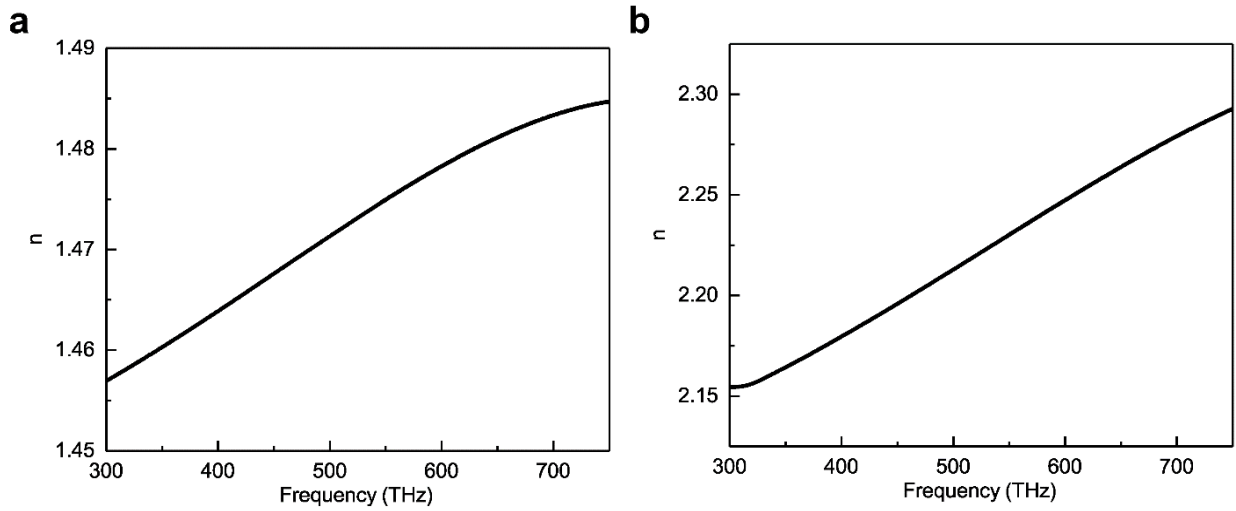

Fig. S9 Measured refractive indexes (*n*) of $SiO_2$ (**a**) and $Ta_2O_5$ (**b**) for the frequencies of interest. $SiO_2$ and $Ta_2O_5$ are nonmagnetic material with relative permeability both equal to 1.

**Supplementary References**

1. Xiao, M., Zhang, Z. Q. & Chan, C. T. Surface impedance and bulk band geometric phases in one-dimensional systems. *Physical Review X* **4**, 021017 (2014).
2. Yariv, A. & Yeh, P. Photonics: Optical Electronics in Modern Communications Ch. 12 (Oxford Univ. Press, Oxford, 2007).